# Superconductivity in $La_{1.85}Sr_{0.15}CuO_4$ Ceramic Samples

N. V. Dalakova[†,1], B. I. Belevtsev[†,1], E. Yu. Belyaev[1], Yu. A. Savina[1], O. I. Yuzephovich[†,1,2], S. V. Bengus[1,2], N. P. Bobrysheva[3]

[1] B. Verkin Institute for Low Temperature Physics and Engineering, National Academy of Sciences of Ukraine, Kharkiv, Ukraine
[2] International Laboratory of High Magnetic Fields and Low Temperatures, Wrocław, Poland
[3] Department of Chemistry, Saint Petersburg State University, Russia

E-mail: beliayev@ilt.kharkov.ua

Effects related to the granularity of a $La_{1.85}Sr_{0.15}CuO_4$ ceramic sample, synthesized by the solid-state reaction method, are presented. The superconducting transition exhibits a step-like behavior. Low-temperature features of magnetoresistance hysteresis loops associated with the granular structure of the sample have been observed.

Polycrystalline (ceramic) high-temperature superconducting (HTSC) cuprates with grain sizes of several microns can be considered as ensembles of type-II superconducting grains. The macroscopic superconducting behavior of such materials is determined by weak links between grains. Typically, each grain is connected to several neighboring grains, forming a three-dimensional network of Josephson junctions [1]. These intergranular connections may significantly vary in their nature and properties [2].

For this reason, in granular superconductors, the resistive superconducting transition is notably broader than in homogeneous systems, and a characteristic feature of such materials is a two-step transition. As temperature decreases, a marked drop in resistance occurs at the onset temperature $T_{co}$, associated with superconducting transitions within individual grains. However, due to weak intergranular coupling, the overall resistance does not immediately vanish. The further evolution of the resistance $R(T)$ below $T_{co}$ depends on the strength and distribution of intergranular coupling [3]. With decreasing temperature, the Josephson coupling strengthens, and discontinuous zero-resistance pathways or clusters begin to emerge, leading to a further decline in resistance.

Due to inevitable variations in grain boundary thickness – and, consequently, in contact resistance – the resistive transition acquires a percolative character. With lowering temperature, superconducting clusters expand, and, provided the spatial disorder in contact resistances is not too strong, continuous percolating superconducting pathways emerge at a critical temperature $T_c$, at which the resistance drops to zero. This global superconducting $T_c$ can be significantly lower than the intragrain onset temperature $T_{co}$.

In such granular systems, the mechanism of magnetic flux trapping near the superconducting transition remains not fully understood. It is still unclear whether the flux is pinned within individual grains or within closed superconducting loops formed by the percolative network.

In this study, we investigated the transport and magnetotransport properties of a $La_{1.85}Sr_{0.15}CuO_4$ cuprate sample synthesized via the standard solid-state reaction method. The sample was characterized using X-ray diffraction, magnetic measurements, electron microscopy, and complementary techniques.

The microstructure, elemental composition, and phase distribution were examined by scanning electron microscopy (SEM) using a CamScan instrument. The concentrations of copper and lanthanum were determined by energy-dispersive X-ray spectroscopy (EDS, LINK AN–10000), while the strontium content was measured using a high-sensitivity wavelength-dispersive spectrometer (WDS, MIKROSPEC) at five different locations across the sample. The elemental ratios were consistent with the nominal chemical formula. The grain size of the ceramic sample was approximately 1–3 μm.

Fig. 1 shows the temperature dependence of the magnetic susceptibility for the investigated sample, measured using a SQUID magnetometer (MPMS–XL5). The superconducting transition temperature $T_{co1}$, determined by the onset of the diamagnetic response in a magnetic field of 10 Oe, was found to be 35.6 K.

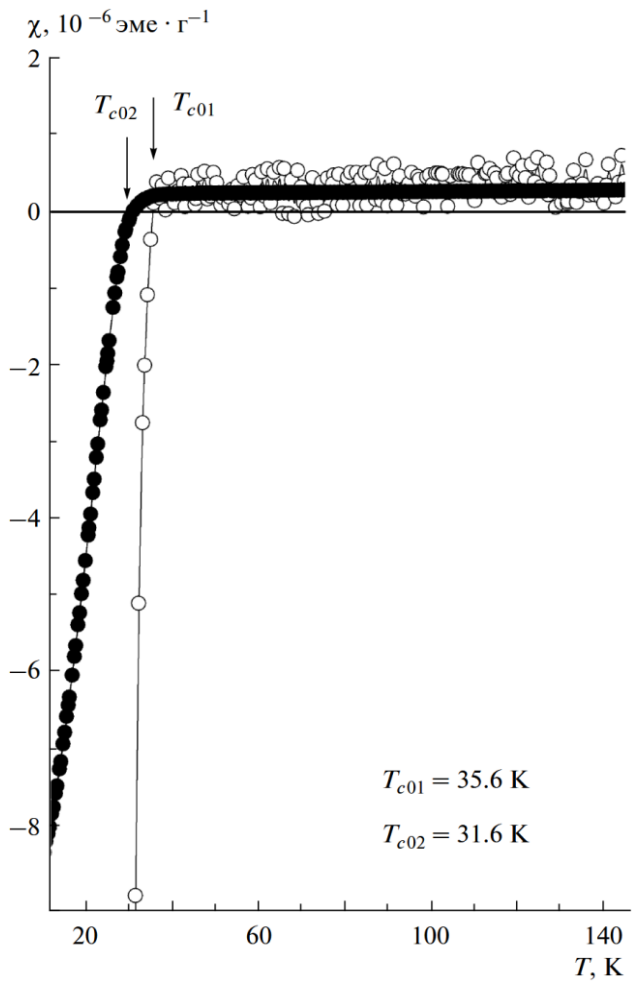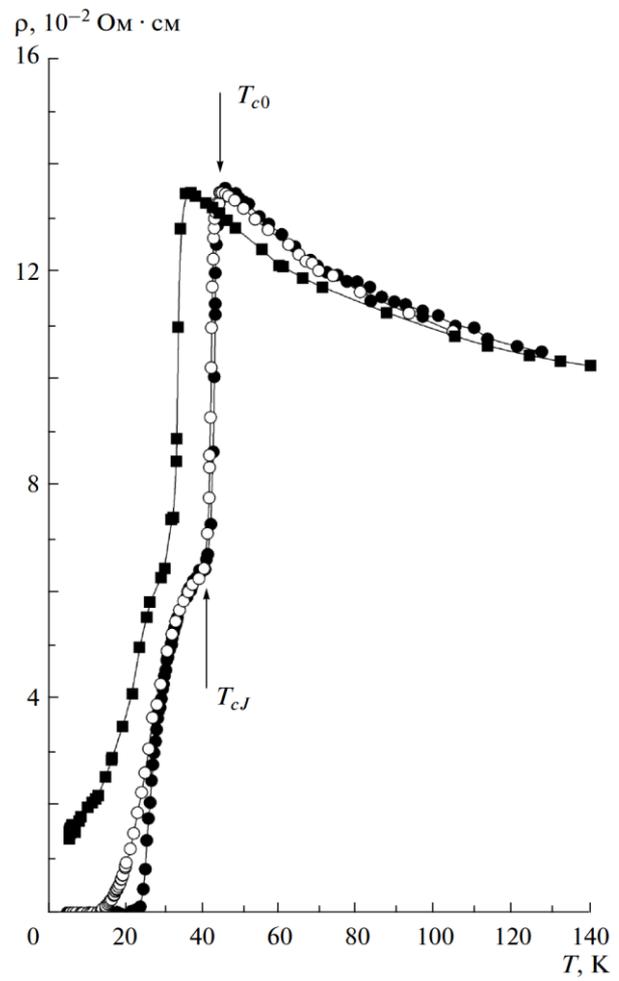

Fig. 1. Temperature dependence of the magnetic susceptibility of the $La_{1.85}Sr_{0.15}CuO_4$ sample. Open symbols correspond to $H = 10$ Oe; filled symbols to $H = 10$ kOe.

Figure 2. Temperature dependence of resistivity for the $La_{1.85}Sr_{0.15}CuO_4$ sample measured at different current levels: ▲ – 0.3 mA; ○ – 3 mA; ■ – 10 mA.

The temperature dependence of resistivity was measured at different current levels in the range of 0.01–10 mA. The ρ(T) curves recorded at low currents (0.01–0.3 mA) coincide in the region of the superconducting transition. The onset temperature for the intergranular superconducting transition, $T_{co} = 44$ K (indicated by the arrow in Fig. 2), as determined from the condition $d\rho/dT = 0$ at $J = 300$ μA, marks the establishment of superconductivity in the percolative network being formed by the grain ensemble.

Above $T_{co}$, the non-metallic behavior of R(T) ($d\rho/dT < 0$) reflected weak intergranular coupling. Between $T_{co}$ and 200 K, the resistivity followed an approximately logarithmic temperature dependence, while above 200 K, a metallic conduction regime set in.

Below $T_{co}$, the resistivity decreased rapidly with decreasing temperature, reaching $T_{cJ} \approx 40$ K, where Josephson coupling in disordered media between superconducting grains began to form. In the range ~40–35.6 K, a step-like feature appeared in the resistivity curve, followed by a sharp drop to zero as percolative Josephson paths developed throughout the intergranular network (see Fig. 2). The superconducting transition came to completion at $T \approx 23$ K for $J \le 0.3$ mA.

At the onset temperature of the superconducting transition, $T_{co} = 44$ K, no anomalies were detected in the χ(T) curves. The transition temperature determined from magnetic measurements, $T_{co1} = 35.6$ K, is significantly lower than $T_{cJ}$. In the resistive R(T) curves, a sharp drop in resistance to zero begins precisely at $T_{co1} = 35.6$ K. This indicated that the diamagnetic response (see Fig. 1) emerged only when a sufficiently large volume of the superconducting phase formed in disordered intergranular media with lowered $T_{co1} < T_{cJ}$ and phase coherence was established between superconducting grains with sufficient spatial extent. Thus, the temperature $T_{co1}$ = 35.6 K corresponded to the transition of the system into a globally coherent superconducting state, as it was reflected in its bulk magnetic response.

The resistive superconducting transition manifested a pronounced dependence on both the applied current (Fig. 2) and the external magnetic field (Fig. 3). As the current increased, at $J$ = 3 mA, the intergranular transition became noticeably broadened in the temperature range $T < T_{cJ}$. With further increase in current, some of the weakest links in the percolative paths – those with the lowest critical currents – switched to the resistive state, thereby increasing the overall resistance of the system. As a result, the onset temperature $T_{co}$ shifted significantly downward (see Fig. 2, $J$ = 10 mA).

A similar effect was produced by the magnetic field in the temperature range below $T_{co}$ (Fig. 3). At low fields ($H$ < 0.8 kOe), the magnetic field disrupted phase coherence between individual grains, resulting in a sharp increase in resistance for $T < T_{cJ}$ and a noticeable broadening of the superconducting transition. As the field was increased to $H$ = 10 kOe, intragranular superconductivity began to break down, and the onset temperature of the global superconducting transition gradually decreased.

Thus, relatively weak currents and moderate magnetic fields suppressed the Josephson coupling between grains without significantly affecting intragranular superconductivity. Nevertheless, these small changes had a strong impact on the shape of the resistive transition as a whole. The data presented in Figs. 2 and 3 indicated that the critical fields and currents of the superconducting grains substantially exceeded those of the intergranular medium. The step-like nature of the superconducting transition in inhomogeneous superconductors was discussed in detail in [4].

The study of the magnetoresistive (MR) effect revealed a number of features in the magnetic field dependence of resistance that correlated with the results of resistive measurements. Above $T_{co}$, a negative MR was observed. At $T < T_{co}$, the MR became positive. In the region of the superconducting transition (around $T = T_{cJ}$), a crossover was observed from a counterclockwise MR hysteresis loop (in which $\Delta R(H)$ during decreasing field lies above that during increasing field) to a clockwise hysteresis loop.

At $T < T_{cJ}$, the hysteresis loop was clockwise, and the MR saturated at relatively low magnetic fields. In contrast, for $T > T_{cJ}$, the hysteresis was counterclockwise, and no saturation occurred (Fig. 4). These data correlated with the results shown in Fig. 3, which demonstrated that at fixed temperatures $T < T_{cJ}$, the resistance sharply increased in low fields $H$ < 0.8 kOe and then remained nearly unchanged up to 10 kOe. This behavior indicated that intergranular superconductivity was almost completely suppressed by weak magnetic fields ($H \approx 0.8$ kOe) in the temperature range $\approx$20–35 K (Figs. 3, 4). For $T > T_{cJ}$, the resistance gradually increased with magnetic field up to 140 kOe. In fields $H \leq 10$ kOe, this increase was less pronounced than that observed below $T_{cJ}$.

High-temperature granular superconductors are inherently inhomogeneous systems, and their properties are typically interpreted within the framework of the two-level critical state model [5]. This model describes a two-component system consisting of (i) superconducting grains characterized by a critical current density $J_{cg}$ and lower and upper critical fields $H_{c1g}$ and $H_{c2g}$, and (ii) an intergranular Josephson medium with its own critical parameters: $J_{cJ}$, $H_{c1J}$, and $H_{c2J}$.

In the mixed state of such an inhomogeneous system, the resistive behavior is governed primarily by the intergranular coupling. The Josephson intergranular medium was found to be highly sensitive to small variations in current and magnetic field, whereas such changes (for $J \leq J_{cg}$ and $H \leq H_{c1g}$) had little effect on the intragranular superconducting properties. This type of behavior was clearly demonstrated in Figs. 2 and 3.

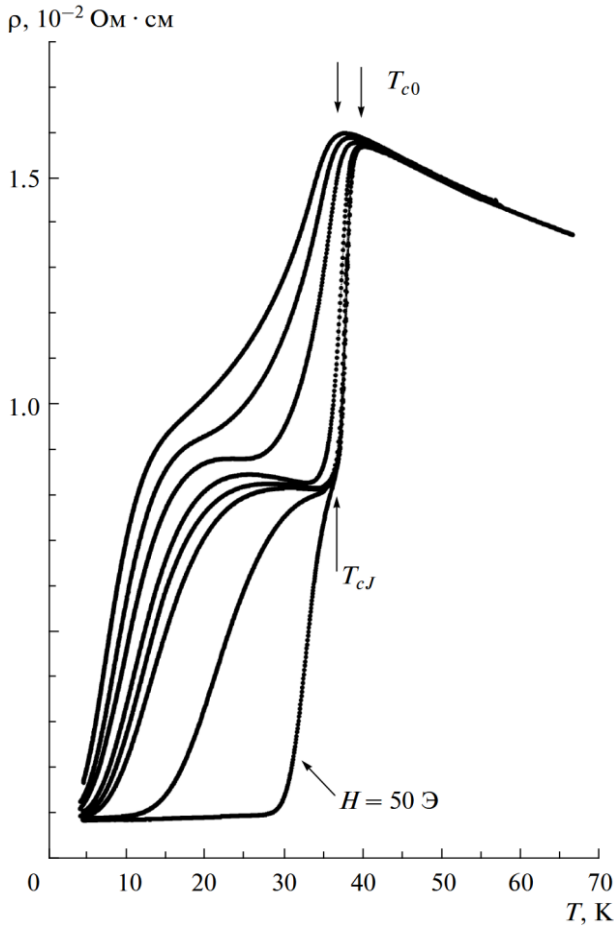
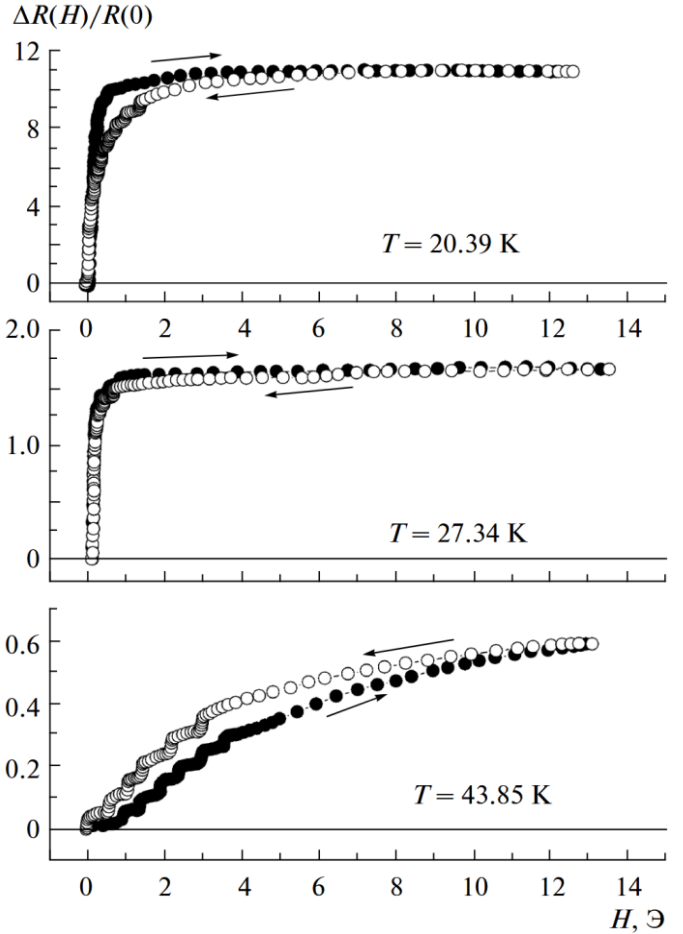

Fig. 3. Temperature dependence of resistivity of the $La_{1.85}Sr_{0.15}CuO_4$ ceramic sample measured under various magnetic fields: $H = 0.05, 0.1, 0.8, 2, 10, 50, 100,$ and $140$ kOe. $\vec{H} \perp \vec{J}$.

Fig. 4. Magnetic field dependence of the resistivity of the $La_{1.85}Sr_{0.15}CuO_4$ ceramic sample measured at three different temperatures and current $J = 300$ μA. $\vec{H} \perp \vec{J}$.

The key parameter governing the field dependence of resistance in granular superconductors is the magnetic flux density (induction) $B_J$ in the intergranular Josephson medium. A clear correlation between $B_J$ and the resistance in the mixed state of granular high-temperature superconductors was demonstrated in [6].

According to the two-level critical state model [5], the intergranular flux density $B_J$ is determined by three contributions: (1) the applied magnetic field $H_{ap}$, (2) the intragranular magnetization $M_g$, and (3) the intergranular magnetization $M_J$. Due to the weakened superconductivity in the intergranular medium, for the temperature range $T_{co} > T > T_{cJ}$ the inequality $M_J \ll M_g$ holds, and thus, to a first approximation, the intergranular induction can be written as $B_J = H_{ap} - M_g(H_{ap}) \times C(H_{ap})$, where $C(H_{ap})$ is a numerical coefficient dependent on the grain shape.

It followed that the MR hysteresis was primarily governed by the hysteretic behaviour of the intragranular magnetization, which is characterized by flux trapping. In superconducting grains, the $M(H_{ap})$ curve during decreasing field lies above that during increasing field, due to the trapped flux. As a result, the total flux density $B_J$ in the granular superconductor was higher in increasing fields (according to the expression for $B_J$), leading to a clockwise hysteresis loop (Fig. 4, $T = 20.39$ K, $T = 27.34$ K).

At temperatures $T \geq T_{cJ}$, the granular system no longer follows the two-level critical state model [5], as the intergranular Josephson medium is no longer present. The system effectively transforms into a single-component medium composed of superconducting grains, and in the temperature interval $T_{cJ} \leq T < T_{co}$, its magnetic superconducting response becomes homogeneous. Accordingly, a counterclockwise hysteresis loop, typical of homogeneous superconductors and associated with flux trapping, is observed in this regime (Fig. 4, $T = 43.85$ K).

## Conclusions

Analyzing the magnetoresistance and magnetization data obtained for a polycrystalline ceramic sample of $La_{1.85}Sr_{0.15}CuO_4$, we have shown that granular HTSC samples represent an inhomogeneous system in which superconducting phase coherence at temperatures $T < T_{cJ}$ is established via Josephson coupling between superconducting grains. A new effect was observed and interpreted: a transition at $T_{cJ}$ from a counterclockwise (for $T > T_{cJ}$) to a clockwise magnetoresistance hysteresis loop (for $T < T_{cJ}$). This transition marks a qualitative change in intergranular superconducting coherence and is consistently explained within the two-level critical state model, demonstrating its applicability to the granular superconducting state in the intermediate regime $T_{cJ} < T < T_{co}$.

Notably, the onset of the superconducting transition as determined from resistivity measurements ($T_{co} \approx 44$ K) did not coincide with the appearance of the diamagnetic response in $\chi(T)$, which occurred only at $T_{co_1} = 35.6$ K. This discrepancy is characteristic of granular superconductors, where intragranular superconductivity precedes the establishment of global phase coherence required for a macroscopic diamagnetic response.

Also, the observed transition in MR loops' hysteresis direction, together with the offset between resistive and magnetic transitions, offers further insight into the long-standing question of where magnetic flux trapping originates in granular superconductors — whether within individual grains or in closed superconducting loops formed by the percolative network. Our data suggest that flux trapping and the associated magnetic response arise only when phase coherence spans a sufficiently large portion of the superconducting percolative cluster being formed in the granular structure, supporting the latter scenario.

### Acknowledgements
E.Yu. Beliayev gratefully acknowledges the financial support provided by the Carlsberg Foundation (Denmark) through the SARU (Scholars at Risk from Ukrainian Universities) Fellowship, which enabled the continuation of his scientific work at the University of Southern Denmark, Sønderborg, during this critical wartime period.

I also wish to express my deep gratitude to my late colleagues and co-authors — Nina Dalakova, Boris Belevtsev, and Olga Yuzephovich — who initiated the original research.

Additionally, I am sincerely grateful to Prof. Yogendra Kumar Mishra and Prof. Horst-Günter Rubahn for their kind invitation and continued support, which made it possible to revisit and expand this earlier work.

This is an author-prepared English version of a 2014 article originally published in Russian (*Izvestiya RAN, Seriya Fizicheskaya*). The text has been extended and restructured in response to several inquiries from colleagues and my reanalysis of the original data. The present version refines the conclusions and contextualizes them within the broader framework of granular superconductivity and the percolation approach.

## References

1. T. P. Sheahen, *Introduction to High-Temperature Superconductivity*, Kluwer Academic, New York, 2002, 580 p.

2. H. Hilgenkamp, J. Mannhart, "Grain boundaries in high-$T_c$ superconductors", *Rev. Mod. Phys.* **74** (2002) 485.

3. B. I. Belevtsev, E. Yu. Beliayev, D. G. Naugle, et al., "Granular behavior of the high-temperature superconductor $La_{2-x}Sr_xCuO_4$", *J. Phys.: Condens. Matter* **19** (2007) 036222.

4. I. M. Dmitrenko, A. M. Glukhov, A. S. Zaika, et al., "Features of the resistive superconducting transition in granular HTSC", *Fiz. Nizk. Temp.* (Low Temp. Phys.) **14** (1988) 1045.

5. L. Ji, M. S. Rzchowski, N. Anand, M. Tinkham, "Magnetoresistance and critical currents in granular $YBa_2Cu_3O_7$", *Phys. Rev. B* **47** (1993) 470.

6. A. I. Kopeliovich, "On the role of demagnetization and magnetization currents in granular superconductors", *Fiz. Tverd. Tela* (Sov. Phys. Solid State) **32** (1990) 3613.